\documentclass[prd,twocolumn,amsfonts,amssymb]{revtex4}

\usepackage{graphicx}

\usepackage{bm}

\newcommand{\beq}{\begin{equation}}
\newcommand{\eeq}{\end{equation}}
\newcommand{\beqs}{\begin{eqnarray}}
\newcommand{\eeqs}{\end{eqnarray}}

\newcommand{\drawsquare}[2]{\hbox{%
\rule{#2pt}{#1pt}\hskip-#2pt
\rule{#1pt}{#2pt}\hskip-#1pt
\rule[#1pt]{#1pt}{#2pt}}\rule[#1pt]{#2pt}{#2pt}\hskip-#2pt
\rule{#2pt}{#1pt}}
\newcommand{\fund}{\raisebox{-.5pt}{\drawsquare{6.5}{0.4}}}

\begin{document}

\title{Analysis of a Zero of a Beta Function Using All-Orders 
Summation of Diagrams}

\author{Robert Shrock}

\affiliation{C.N. Yang Institute for Theoretical Physics, Stony Brook
  University, Stony Brook, NY 11794}

\begin{abstract}

Conventionally, one calculates a zero in a beta function by computing this
function to a given loop order and solving for the zero.  Here we discuss a
different method which is applicable in theories where one can perform a
partial diagrammatic summation to infinite-loop order.  We show that this
method, compared with the conventional method, yields much better agreement
with exact results in the case of an asymptotically free gauge theory with
${\cal N}=1$ supersymmetry.  Applications to other field theories are also
discussed.

\end{abstract}


\maketitle

In a quantum field theory, the dependence of the interaction coupling on the
Euclidean momentum scale, $\mu$, where it is measured is of fundamental
importance. This dependence is described by the beta function, $\beta$, of the
theory \cite{rg}. Of particular interest is a zero of $\beta$, since at this
zero, the coupling is scale-invariant.  We focus on a zero of $\beta$ at
nonvanishing coupling. The (physical) zero of $\beta$ closest to the origin in
coupling constant space is an infrared (IR) zero for an asymptotically free
theory and an ultraviolet (UV) zero for an IR-free theory. In a conventional
analysis of the beta function of a given theory, one considers
the series expansion of $\beta$ to a given order in powers of the coupling and
investigates whether or not it has a (physical) zero away from the origin in
coupling-constant space.

Here we discuss a different approach to the calculation of a zero of the beta
function. Our approach is applicable to theories where one can perform a
summation, to infinite-loop order, of the terms from a (sub)set of diagrams
contributing to the beta function.  Although our method can be applied to any
theory where one can carry out this type of summation, it is conveniently
illustrated in the context of an asymptotically free vectorial gauge theory.
Thus, we consider such a gauge theory, with gauge group $G$, running coupling
$g=g(\mu)$, and $N_f$ massless Dirac fermions in a representation $R$ of $G$
\cite{fm}.  Let $\alpha(\mu) = g(\mu)^2/(4\pi)$.  If $N_f$ is sufficiently
large (still maintaining asymptotic freedom), then the two-loop beta function
has an IR zero.  As $\mu$ decreases from large values in the UV, $\alpha(\mu)$
increases toward the value of this zero, which thus plays an important role in
determining the IR properties of the theory \cite{irz}. The beta function is
$\beta_g = dg/dt$, where $dt=d\ln\mu$, or equivalently, $\beta_\alpha =
d\alpha/dt = [g/(2\pi)]\beta_g$. This has the series expansion
\beq
\beta_\alpha = -2\alpha \, \sum_{\ell=1}^\infty b_\ell \, a^\ell \ , 
\label{beta}
\eeq
where $b_\ell$ is the $\ell$-loop coefficient and $a=g^2/(16\pi^2)$.  The
coefficients $b_1$ and $b_2$ are independent of the regularization scheme,
while the $b_\ell$ for $\ell \ge 3$ are scheme-dependent \cite{gross75}. The
property of asymptotic freedom means $b_1 > 0$. The $n$-loop ($n\ell$) beta
function, denoted $\beta_{\alpha,n\ell}$, is given by Eq. (\ref{beta}) with
$\ell=n$ as the upper limit on the sum.  We define the reduced ($r$) function
\beq
\beta_{\alpha,r} \equiv - \frac{\beta_\alpha }{[\alpha^2 \, b_1/(2\pi)]} = 
1 + \frac{1}{b_1} \, \sum_{\ell=2}^\infty b_\ell \, a^{\ell-1}
\label{betareduced}
\eeq
and $\beta_{\alpha,r,n\ell}$, given by Eq. (\ref{betareduced}) with the upper
limit $\ell=n$ on the sum.  The conventional approach to the study of a
possible IR zero in $\beta_{\alpha,n\ell}$ consists of an analysis of the $n-1$
zeros of the $n$-loop reduced beta function $\beta_{\alpha,r,n\ell}$.  In
particular, at the two-loop level, the IR zero is given by $\alpha_{IR,2\ell} =
- 4\pi b_1/b_2$, which is physical if $b_2 < 0$. Since we are interested in an
IR zero here, we thus assume that the matter content of the theory is such that
$b_2 < 0$.

However, if one can carry out a summation of an infinite (sub)set of diagrams
contributing to the beta function, this yields a result that, although formally
equivalent to (\ref{beta}) when expanded in a series, involves a closed-form
($cf$) functional factor, $f_{cf}$. Normalizing this factor so that $f_{cf}=1$
at $a=0$, we can reexpress $\beta_\alpha$ in this case as
\beq
\beta_\alpha = - \frac{\alpha^2 b_1}{2\pi} \, f_{cf} \, f_s \ , 
\label{betaff}
\eeq
i.e., $\beta_{\alpha,r}=f_{cf} \, f_s$, where $f_s$ is given by a series ($s$)
expansion
\beq
f_s = 1 + \sum_{j=1}^\infty f_{s,j} \, a^j \ .
\label{fs}
\eeq

In a theory where this summation can be carried out and $\beta_\alpha$ can thus
be expressed in the form (\ref{betaff}), our proposed method is to analyze the
zeros in $f_{cf}$ and $f_s$ rather than the zeros of $\beta_{\alpha,r,n\ell}$
as a means of gaining information about the physical zero in the beta function.
This method takes advantage of the information from the all-orders summation,
while the conventional method of analyzing zeros in $\beta_{\alpha,r,n\ell}$ to
a given finite loop order $n$ does not. In particular, if $f_{cf}$ has no
physical zero, then the proposed method is to examine the zeros in $f_s$. 

Let us consider an illustrative function for $f_{cf}$, namely 
\beq
f_{cf} = \frac{1}{1+d_1 a} \ , 
\label{fcf}
\eeq
(where $d_1$ is a constant), so that $\beta_{\alpha,r}=f_s/(1+d_1 \, a)$. 
Then
\beq
f_s = 1 + \frac{1}{b_1} \, \sum_{n=2}^\infty 
( b_n + d_1 \, b_{n-1}) \, a^{n-1} \ . 
\label{fsd1}
\eeq
Solving the equation $f_s=0$ to the lowest nontrivial order, $n=2$, we obtain
for the IR zero
\beq
\alpha_{IR,cfs}= -\frac{4\pi}{ (b_2/b_1) + d_1} \ , 
\label{alfir_cfs2}
\eeq
which is physical if $(b_2/b_1) + d_1 < 0$ (the subscript $cfs$ connotes the
direct dependence on $f_s$ and the implicit dependence on $f_{cf}$.)  Since we
assume $b_2 < 0$ (and $b_1 > 0$), $\alpha_{IR,cfs}$ is physical if
\beq
d_1 < \frac{|b_2|}{b_1} \ . 
\label{d1ineq}
\eeq
Because the existence of the IR zero in $\beta_{\alpha,IR,2\ell}$
is scheme-independent, one may require that the scheme used for the summation
to infinite-loop order that yields (\ref{betaff})
should not remove this IR zero, and hence that $d_1$ satisfies the inequality
(\ref{d1ineq}). As compared with $\alpha_{IR,2\ell}$, 
\beq
\alpha_{IR,cfs} = \frac{\alpha_{IR,2\ell}}
{1 - d_1 \, \frac{a_{IR,2\ell}}{4\pi} } \ . 
\label{alf_cfs2}
\eeq
Hence, if $d_1$ is negative (positive), then $\alpha_{IR,cfs}$ is smaller
(larger) than $\alpha_{IR,2\ell}$.  As is evident from the dependence on
$d_1$, $\alpha_{IR,cfs}$ incorporates information from the summation to
infinite-loop order, in contrast to $a_{IR,n\ell}$ for any finite $n$. 
A necessary condition for the result
(\ref{alf_cfs2}) to be physically meaningful is that if $f_{cf}$ has any
divergent physical singularities, they must occur farther from the origin than
$\alpha_{IR,cfs}$.  In our example, $f_{cf}$ has a simple pole at
$\alpha_{cf,pole}=-4\pi/d_1$, which occurs at physical coupling if $d_1 < 0$.
The required inequality is $\alpha_{cf,pole} > \alpha_{IR,cfs}$, i.e., 
\beq
-\frac{1}{d_1} > -\frac{1}{(b_2/b_1)+d_1} \ .
\label{ineq1}
\eeq
Since $b_2 < 0$ for the existence of $\alpha_{IR,2\ell}$ and we only need to
address the case where $d_1$ is negative, (\ref{ineq1}) is equivalent to the
inequality
\beq
\frac{1}{|d_1|} > \frac{1}{(|b_2|/b_1)+|d_1|} \ ,
\label{ineq11}
\eeq
which is clearly true.  This proves that (for the case we consider here, where
the theory has an IR zero in $\beta_{\alpha,2\ell}$) even if $f_{cf}$ contains
a physical pole, one has $\alpha_{cf,pole} > \alpha_{IR,cfs}$, so this pole
is irrelevant for the evolution from weak coupling in the UV to the IR zero at
$\alpha_{IR,cfs}$. 

To demonstrate the usefulness of the proposed method, we apply it to an
asymptotically free vectorial ${\cal N}=1$ supersymmetric gauge theory (denoted
SGT) with gauge group SU($N_c$) and $N_f$ pairs of massless chiral superfields
$\Phi_j$, $\tilde \Phi_j$, $1 \le j \le N_f$ transforming according to the
representations $R$ and $\bar R$ of SU($N_c$), respectively. In this theory,
the scheme-independent coefficients in $\beta_\alpha$ are $b_1 = 3C_A - 2T_f
N_f$ and $b_2 = 6C_A^2-4(C_A+2C_f)T_fN_f$ \cite{b12susy,casimir}.  Calculations
to three-loop \cite{b3susy} order have been done, e.g., in the $\overline{\rm
  DR}$ scheme \cite{drbar}. This theory has the appeal that a number of exact
results are known for it \cite{nsvz}-\cite{seiberg}, enabling one to compare
approximate calculations with these exact results \cite{sd}-\cite{lnn}. We will
show that the method discussed here, compared with the conventional method,
yields much better agreement with exact results and solves a puzzle found in
\cite{bfs}.

In this SGT the constraint of asymptotic freedom requires that $N_f <
N_{f,b1z}$, where $N_{f,b1z}=3C_A/(2T_f)$ \cite{nfintegral}. It is
known \cite{nsvz}-\cite{seiberg} that if $N_f$ is in the interval $I_{NACP}$ defined by $N_{f,cr} < N_f
< N_{f,b1z}$, where
\beq
N_{f,cr} = \frac{3C_A}{4T_f} = \frac{N_{f,b1z}}{2} \ ,
\label{nfcr}
\eeq
then this theory flows from weak coupling in the UV to a conformal non-Abelian
Couloumb phase (NACP) in the IR without any spontantaneous chiral symmetry
breaking (S$\chi$SB). A quantity of interest is the anomalous dimension of the
fermion bilinear, $\gamma_m$ \cite{gammadef}. This has the series expansion
\beq
\gamma_m = \sum_{\ell=1}^\infty c_\ell a^\ell \ , 
\label{gamma_m}
\eeq
where $c_1=4C_f$ is scheme-independent, and the $c_\ell$ with $\ell \ge 2$ are
scheme-dependent.  Using instanton methods, Novikov, Shifman, Vainshtein, and
Zakharov (NSVZ) obtained an exact closed-form solution for $\beta_\alpha$
\cite{nsvz,sv} (in a particular scheme \cite{schd}):
\beq
\beta_{\alpha,NSVZ} = - \frac{\alpha^2}{2\pi} \, \bigg [ 
\frac{b_1-2T_fN_f\gamma_m }{1- 2C_A \, a } \bigg ] \ .
\label{beta_nsvz}
\eeq
As $N_f$ decreases from $N_{f,b1z}$ and approaches
$N_{f,cr}$ from above, $\gamma_m$ increases from 0 to the maximum value allowed
by unitarity in this conformal phase, namely $\gamma_m=1$
\cite{gammabound,mack_etal}.  Applying the condition that
$\gamma_m$ should saturate this unitarity upper bound at an IR zero of $\beta$
as $N_f$ approaches $N_{f,cr}$ from above is one way to derive the value of
$N_{f,cr}$. 

Using known results for the $b_\ell$ and $c_\ell$ for $1 \le \ell \le 3$
\cite{b12susy}-\cite{b3susy}, Ref. \cite{bfs} calculated $\alpha_{IR,n\ell}$ at
the $n=2$ and $n=3$ loop level and evaluated the $n$-loop anomalous dimension
of the fermion bilinear at $\alpha=\alpha_{IR,n\ell}$, denoted
$\gamma_{IR,n\ell}$.  Although higher-order calculations of $\alpha_{IR,n\ell}$
are scheme-dependent, they are quite valuable in obtaining more accurate
information about $\gamma_m$ at an exact IR fixed point, which is a universal
quantity.  Ref. \cite{bfs} found that at the two-loop level,
$\gamma_{IR,2\ell}$ increases monotonically as $N_f$ decreases from
$N_{f,b1z}$, but exceeds unity at a value of $N_f \in I_{NACP}$ larger than
$N_{f,cr}$. As was noted in \cite{bfs}, this is understandable in view of the
fact that the two-loop perturbative calculations of $\beta$ and $\gamma_m$ are
not expected to agree precisely with exact results; the differences provide a
quantitative measure of the accuracy of these two-loop calculations.  Normally,
one would expect that calculating a quantity to higher-loop order should give a
more accurate result if perturbative computations are reliable. However, the
study of the IR behavior at three-loop level in \cite{bfs} yielded a different
and very puzzling result: $\gamma_{IR,3\ell}$ does not approach more closely to
1 as $N_f$ decreases through $I_{NACP}$ toward $N_{f,cr}$; on the contrary,
$\gamma_{IR,3\ell}$ reaches a maximum at a small positive value and then
decreases, passing through zero to negative values. 

We next show how our proposed method removes this puzzling behavior and
produces excellent agreement with exact results for this SGT.  First, observe
that $\beta_{\alpha,NSVZ}$ has the form of Eq. (\ref{betaff}) with
\beq
f_{cf,NSVZ} = \frac{1}{1-2C_A \, a } \ , 
\label{fcf_nsvz}
\eeq
and 
\beq
f_{s,NSVZ} = 1-\frac{2T_fN_f \, \gamma_m}{b_1} \ , 
\label{fs_nsvz}
\eeq
We focus on the range of $N_f$ where $\beta_{\alpha,NSVZ}$ has an IR zero. 
Applying our method, we observe that $f_{cf,NSVZ}$ has no zero, so we solve the
equation $f_{s,NSVZ}=0$.  Retaining the maximal scheme-independent information
in $\gamma_m$, namely the $\ell=1$ term in (\ref{gamma_m}), this equation reads
$b_1-8N_f T_f C_f a=0$, whence
\beq
\alpha_{IR,cfs} = \frac{\pi(3C_A-2T_fN_f)}{2T_fC_fN_f} = 
\frac{\pi}{C_f} \, \Big [ \frac{2N_{f,cr}}{N_f}-1 \Big ] \ .
\label{alfir_cfs_nsvz}
\eeq
As $N_f$ decreases from $N_{f,b1z}$ to $N_{f,cr}$, $\alpha_{IR,cfs}$ increases
monotonically from 0 to $\alpha_{cfs}=\pi/C_f$.  Since $d_1 < 0$ (and with $b_2
< 0$ so that $\alpha_{IR,2\ell}$ is physical), it follows, as a special case of
(\ref{alf_cfs2}), that, $\alpha_{IR,cfs} < \alpha_{IR,2\ell}$ in this SGT.  Our
general discussion above shows that the pole in $f_{cf,NSVZ}$ at
$\alpha_{cf,pole,NSVZ}=2\pi/C_A$ lies farther from the origin than
$\alpha_{IR,cfs}$ \cite{nsvzpole}. As $N_f$ approaches $N_{f,b1z}$ from below,
both $\alpha_{IR,cfs}$ and $\alpha_{IR,2\ell}$ approach zero, and the ratio
$\alpha_{IR,cfs}/\alpha_{IR,2\ell} \nearrow 1$.

Thus, $\alpha_{IR,cfs}$ retains perturbative reliability much better than the
conventional two-loop result
\beq
\alpha_{IR,2\ell} = \frac{2\pi(3C_A-2T_fN_f)}{2(C_A+2C_f)T_fN_f-3C_A^2} \ , 
\label{alfir2loop}
\eeq
which diverges as $N_f \searrow N_{f,b2z}=3C_A^2/[2(C_A+2C_f)T_f]$, where 
$b_2 \to 0$.  For $R$ equal to the fundamental representation (with Young 
tableau $\fund$), $N_{f,b2z} \in I_{NACP}$.

Next, we study $\gamma_m$, calculated to its maximal scheme-independent
order, $\gamma_m = 4C_f \, a$, and evaluated at $\alpha_{IR,cfs}$. We obtain 
\beq
\gamma_{IR,cfs} = \frac{3C_A}{2T_fN_f}-1 = \frac{N_{f,b1z}}{N_f}-1 = 
\frac{2N_{f,cr}}{N_f}-1 \ .
\label{gamma_ir_cfs_nsvz}
\eeq
From (\ref{gamma_ir_cfs_nsvz}), it follows that $\gamma_m \searrow 0$ as $N_f
\nearrow N_{f,b1z}$.  Furthermore, from (\ref{gamma_ir_cfs_nsvz}) we obtain
three important results for $\gamma_m$, which exhibit complete agreement with
exact results and avoid the problems with $\gamma_{IR,2\ell}$ and
$\gamma_{IR,3\ell}$ found in \cite{bfs}. First, Eq. (\ref{gamma_ir_cfs_nsvz})
shows that as $N_f$ decreases from $N_{f,b1z}$ to $N_{f,cr}$ in the interval
$I_{NACP}$, $\gamma_{IR,cfs}$ increases monotonically. In contrast,
$\gamma_{IR,3\ell}$ was found not to be a monotonically increasing function
with decreasing $N_f \in I_{NACP}$ \cite{bfs}.

Second,$\gamma_{IR,cfs} \le 1$ for all $N_f \in I_{NACP}$, in agreement with
the constraint from unitarity. In contrast, $\gamma_{IR,2\ell}=$ was found to
violate this unitarity constraint for a range of $N_f$ toward the lower end of
the interval $I_{NACP}$ \cite{bfs}; indeed, it was found that
$\gamma_{IR,2\ell}$ diverges, just as $\alpha_{IR,2\ell}$ does, as $N_f
\searrow N_{f,b2z}$ \cite{bfs}. As was noted in \cite{bfs}, this
behavior indicated that the perturbative calculations of $\alpha_{IR,2\ell}$
and the resultant evaluation of $\gamma_m$ at this value of $\alpha_{IR,2\ell}$
are not reliable in this region.  With our different procedure, we avoid this
pathological behavior.

Third, we find that 
\beq
\gamma_{IR,cfs} \nearrow 1 \ {\rm as} \ N_f \searrow N_{f,cr} \ . 
\label{gamma_ir_cfs_nsvz_to_1}
\eeq
Thus, this $\gamma_{IR,cfs}$ has precisely the three
properties of the exact all-orders calculation of $\gamma_{IR}$.  Therefore, we
have shown that our method of calculating the IR zero of the
beta function as a zero of $f_s$ rather than as the zero of
$\beta_{\alpha,r,n\ell}$, together with the evaluation of $\gamma_m$ calculated
to its maximal scheme-independent order to obtain $\gamma_{IR,cfs}$, yields
excellent agreement with exact results and avoids the problems with
$\gamma_{IR,2\ell}$ and $\gamma_{IR,3\ell}$ found in \cite{bfs}.  

For the case $R=\fund$, 
Eqs. (\ref{alfir_cfs_nsvz}) and (\ref{gamma_ir_cfs_nsvz}) are
\beq
\alpha_{IR,cfs,\fund} = \frac{2\pi N_c (3N_c-N_f)}{(N_c^2-1)N_f} 
\label{alfir_cfs_nsvz_fund}
\eeq
and
\beq
\gamma_{IR,cfs,\fund} = \frac{3N_c}{N_f}-1 \ . 
\label{gammair_cfs_nsvz_fund}
\eeq
In this $R=\fund$ case, it is also of interest to consider the 
't~Hooft-Veneziano limit $N_c \to \infty$ and $N_f \to \infty$ with $r=N_f/N_c$
fixed and finite, and $\alpha(\mu) N_c$ a finite function of $\mu$. 
In this limit, $I_{NACP}$ is
$3/2 \le r \le 3$.  In terms of $\xi \equiv \alpha N_c$ in this limit, we have
\beq
\xi_{IR,cfs} = \frac{2\pi(3-r)}{r} \ , \quad 
\gamma_{IR,cfs} = \frac{3-r}{r} \ . 
\label{xiir_lnn}
\eeq
In contrast,
\beq
\xi_{IR,2\ell}= \frac{2\pi(3-r)}{2r-3} \ , 
\label{xi_2loop}
\eeq
which diverges as $r \searrow 3/2$ at the lower boundary of $I_{NACP}$, and 
\beq
\gamma_{IR,2\ell}= \frac{r(r-1)(3-r)}{2(3-2r)^2} \ , 
\label{gammair_2loop_lnn}
\eeq
which violates its unitarity upper bound as $r$ decreases through $r=2$ in the
interior of $I_{NACP}$ \cite{bfs,lnn} and diverges as $r \searrow 3/2$.

Since the $b_\ell$ with $\ell \ge 3$ are scheme-dependent, so is the
closed-form function $f_{cf}$ resulting from the infinite-loop summation.  One
might think that the analysis of an IR zero in $\beta_\alpha$ could be
simplified by applying a scheme transformation (ST) to eliminate the $b_\ell$
with $\ell \ge 3$. An explicit ST that reduces $\beta_\alpha$ to the two-loop
expression in the local neighborhood of the origin, $\alpha=0$, was constructed
and studied in \cite{scc}-\cite{sch23} (see also \cite{schh}).  However, it was
shown that STs that are acceptable for small $\alpha$ often lead to unphysical
results when applied at a generic zero of a beta function away from the origin
\cite{scc}-\cite{sch23}. We find this to be true for STs that attempt to remove
$b_\ell$ with $\ell \ge 3$ in the present SGT. Hence, one must deal with the
full series (\ref{beta}), where the method discussed here greatly improves the
analysis. 

Another demonstration of the usefulness of the method discussed here is
obtained from analysis of a test function for $\beta_\alpha$. A basic question
for a study of the zero(s) of a beta function and, more generally, in
mathematics is: how well do the zero(s) of a truncation of a power series for a
function $f(z)$ reproduce the zero(s) of the exact function. In Ref. \cite{sch}
this question was investigated for a general asymptotically free gauge theory,
using for $\beta_{\alpha,r}$ the test function
\beq
\beta_{\alpha,r} = h(\alpha) = \frac{\sin (\pi \sqrt{\tilde\alpha} \ ) }
{( \pi \sqrt{\tilde\alpha} \ ) } \ ,
\label{h}
\eeq
where $\tilde\alpha \equiv \alpha/\alpha_{IR}$. We analyzed the zeros of
truncations of the Taylor series expansion of $h(\alpha)$ to high order and
determined how rapidly these approach $\alpha_{IR}$, i.e., how rapidly the
zeros in the ratios $\tilde \alpha$ approach 1. Here, to apply our method, we
start from the above series for $h(\alpha)$ and perform an infinite-order
summation of part of it, using the identity $(\sin x)/x =
[(2/x)\sin(x/2)]\cos(x/2)$ with $x=\pi\sqrt{\tilde \alpha}$, taking
$(2/x)\sin(x/2)=f_{cf}$ to be the closed-form function and taking
$\cos(x/2)=f_s$, expanded as a series.  That is, we write
\beq
\frac{\sin x}{x} = \sum_{n=0}^\infty \frac{(-1)^n \, x^{2n}}{(2n+1)!} = 
\bigg [ \frac{\sin(x/2)}{x/2} \bigg ] \, \sum_{n=0}^\infty
\frac{(-1)^n \, (x/2)^{2n}}{(2n)!} 
\label{sincos}
\eeq
This $f_{cf}$ has no zero in the interval $0 \le x \le \pi$, i.e., $0 \le
\alpha \le \alpha_{IR}$, so the IR zero in $\beta_{\alpha,r}$ comes from $f_s$.
We calculate this from the truncation of the series for $f_s$ and denote it as
$\alpha_{cfs,n}$.  In Table \ref{alftilde} we list the values of $\delta_n
\equiv \tilde \alpha_{IR,n}-1$ obtained by (i) the conventional method of
analyzing $\beta_{\alpha,r,n}$ as a series, as in \cite{sch}, denoted
$(\delta_n)_h$, and by (ii) the method discussed here, with
$(\delta_n)_{cfs}=\tilde\alpha_{cfs,n}-1$. As is evident, our present
method yields faster convergence toward the exact result.

\begin{table}
\caption{\footnotesize{Values of $(\delta_n)_h$ and $(\delta_n)_{cfs}$ as a
    function of respective series truncation order $n$.}}
\begin{center}
\begin{tabular}{|c|c|c|} \hline\hline
$n$ & $(\delta_n)_h$          & $(\delta_n)_{cfs}$            \\ \hline
 2  & $-0.392$                 & $-0.189$                 \\
 3  & complex                  & $2.78 \times 10^{-2}$    \\
 4  & $-3.97 \times 10^{-2}$   & $-1.13 \times 10^{-3}$   \\
 5  & $4.52 \times 10^{-3}$    & $3.15 \times 10^{-5}$    \\
 6  & $-2.83 \times 10^{-4}$   & $-0.592 \times 10^{-6}$  \\
 7  & $1.35 \times 10^{-5}$    & $0.805 \times 10^{-8}$   \\
 8  & $-0.493 \times 10^{-6}$  & $-0.829 \times 10^{-10}$ \\
\hline\hline
\end{tabular}
\end{center}
\label{alftilde}
\end{table}

For a non-supersymmetric gauge theory (in $d=4$ dimensions) with $N_f$
fermions in a general representation $R$ of the gauge group $G$, a rigorous
closed-form expression for $\beta_\alpha$ has not, so far, been obtained,
although a conjecture for an all-orders beta function has been presented in
\cite{nsc}. Higher-loop analyses of an IR zero in $\beta_\alpha$ have been
performed using the conventional method of studying the zeros of the polynomial
$\beta_{\alpha,r,n\ell}$ up to the four-loop level \cite{gk,bvh}. These
obtained the expected result that calculating $\alpha_{IR,n\ell}$ and
$\gamma_{IR,n\ell}$ to higher-loop order led to more stable results.  For
example, for $G={\rm SU}(3)$ and $N_f=12$ fermions in $R=\fund$,
$\gamma_{IR,2\ell}=0.77$, $\gamma_{IR,3\ell}=0.31$ and $\gamma_{IR,4\ell}=0.25$
\cite{bvh}. The four=loop result agrees well with the lattice measurements
$\gamma=0.27(3)$ and $\gamma_{IR} \simeq 0.25$ \cite{hasenfratz} and
$\gamma=0.235(46)$ \cite{lombardo}\cite{gammalgt}.

There are also IR-free theories where one can obtain an exact closed-form
result for the beta function with a UV zero at nonzero coupling. A first
example is the $N \to \infty$ limit of the nonlinear O($N$) $\sigma$ model in
$d=2+\epsilon$ dimensions\cite{nlsm}. Here, for small $\epsilon$, one finds the
closed-form result \cite{nlsm}
\beq
\beta_x = \frac{dx}{dt} = \epsilon \, x \Big ( 1 - \frac{x}{x_c} \Big ) \ , 
\label{betanlsm}
\eeq
where $\lambda$ is the effective coupling, $\lambda_c = 2\pi\epsilon/N$;
$x=\lim_{N \to \infty} \lambda N$, and $x_c = 2\pi\epsilon$ is the UV zero of
$\beta_x$.  In the notation of Eq. (\ref{betaff}), here $f_{cf} = 1-(x/x_c)$
and $f_s=1$. A second example is supersymmetric quantum electrodynamics (SQED)
with $N_f$ pairs of chiral superfields with charges that can be taken to be 1
and $-1$.  As is well known, $\beta_\alpha$ for this theory can be obtained
from (\ref{beta_nsvz}) by setting $C_A=0$, $T_f=1$, and $C_f=1$, and is
\cite{sqed} $\beta_{\alpha,SQED} = [\alpha^2 \, N_f/(2\pi)] \, [ 1 +
  (\alpha/\pi)]$.  Up to the reversal of sign, this has the form of
Eq. (\ref{betaff}) with $f_{cf} = 1+(\alpha/\pi)$ and and $f_s=1$.  It has no
UV zero. Third, one may consider the SGT with $N_f > N_{f,b1z}$. Using
(\ref{beta_nsvz}) again for this theory, one observes that the numerator of
(\ref{beta_nsvz}) is $-|b_1|-2N_fT_f\gamma_m$. Evaluating $\gamma_m$ at the
maximal scheme-independent level, $\gamma_m=4C_fa$, it follows that this
numerator has no UV zero. Some related work for a nonsupersymmetric U(1) gauge
theory (in $d=4$) is in \cite{holdom,lnf}; in this theory, both a conventional
analysis of $\beta_{\alpha,r}$ up to the five-loop level and use of some
partial closed-form results also led to evidence against a UV zero \cite{lnf}.

In summary, we have discussed a method for calculating a zero in the beta
function of a quantum field theory that is applicable in cases where one can
perform an infinite-loop summation of a subset of diagrams. We have shown that
this method, compared with the conventional fixed-loop-order method, yields
much better agreement with exact results in the case of an asymptotically
free gauge theory with ${\cal N}=1$ supersymmetry.  We have also commented on
the application of this method to some IR-free theories.

This research was partially supported by the grant NSF-PHY-13-16617. I am
grateful to T. A. Ryttov for collaboration on \cite{bfs} and related works. 



\begin{thebibliography}{99}
\bibitem{fm}
%
The generalization to massive fermions is straightforward. As the scale
$\mu$ decreases below the fermion mass $m$, one integrates this fermion out of
the low-energy effective field theory that applies for $\mu < m$.
\bibitem{casimir}
%
The Casimir invariants $C_R$ and $T_R$ are defined in the standard way as
$\sum_a \sum_j {\cal D}_R(T_a)_{ij} {\cal D}_R(T_a)_{jk} = C_R \delta_{ik}$ and
$\sum_{i,j} {\cal D}_R(T_a)_{ij} {\cal D}_R(T_b)_{ji} = T_R \delta_{ab}$, where
$R$ is the representation with matrix ${\cal D}_R(T_a)$, and $T_a$ are the
generators of $G$, so that for SU($N_c$), $C_A=N_c$ for the adjoint ($A$) and
$T_{fund}=1/2$ for the fundamental representation, etc. $C_f$ denotes $C_R$ for
the fermion representation.
\bibitem{nfintegral}
%
Here and below, when an expression such as $N_{f,b1z}$ is given for $N_f$ that
sometimes evaluates to a non-integral real value, it is understood implicitly
that one infers an appropriate integral value of $N_f$ from it.
\bibitem{gammadef}
%
Writing $\Phi_j=\phi_j + \sqrt{2} \, \theta \psi_j+\theta\theta \, F_j$, where
$\theta$ is a Grassmann variable and $F_j$ an auxiliary field, and similarly
with $\tilde \Phi_j$, we consider the holomorphic operator product ${\cal O} =
\Phi_j \tilde \Phi_j$ (no sum on $j$). Denoting the full operator dimension of
${\cal O}$ as $D_{\cal O}$, we have $D_{\cal O}=2-\gamma_m$, where $\gamma_m$
is the anomalous dimension of $\tilde \psi \psi$.
\bibitem{gammabound}
%
In a conformally invariant field theory (whether supersymmetric or not), the
dimension $D_{\cal O}$ of a spin-0 operator ${\cal O}$ (other than the
identity) must be $\ge 1$ to avoid negative-norm states \cite{mack_etal}. This
is equivalent to $\gamma_m \le 1$.
\end{thebibliography}
\end{document}